# Dielectric spheres with maximum forward scattering and zero backscattering:

# A search for their material composition

Yan Zhang,[1] Manuel Nieto-Vesperinas,[2,*] and Juan José Sáenz[1,3,*]

[1]*Condensed Matter Physics Center (IFIMAC), Departamento de Física de la Materia Condensada and Instituto "Nicolás Cabrera", Universidad Autónoma de Madrid, 28049 Madrid, Spain*
[2]*Instituto de Ciencia de Materiales de Madrid, C.S.I.C., Campus de Cantoblanco, 28049 Madrid, Spain*
[3]*Donostia International Physics Center (DIPC),Paseo Manuel Lardizabal 4, 20018 Donostia-San Sebastian, Spain*
*\*Corresponding authors: mnieto@icmm.csic.es, juanjo.saenz@uam.es*

Nanoparticles exhibiting zero backscattering but a large scattering cross section in the forward direction should play a key role as light diffracting elements in photonic devices like solar cells. Using Mie theory we address lossless dielectric spheres that were recently reported to possess a magnetodielectric response to the illuminating wave, and analyze their scattering cross section together with their zero-backwards scattering conditions. We show that there is an optimum particle refractive index ($m$ = 2.47), which yields maximum forward scattering without backwards scattering of light.

In the design of photonic devices such as photodetectors or solar cells, maximum transmission of light into the receiver is desired in order to store as much radiation energy as possible [1-14]. To this end, composite structures made of nano-objects with maximum scattering cross section in the forward half space are worth being analyzed.

In this letter we discuss a new concept that makes use of the condition of zero backward-scattering discussed by Kerker *et al.* [15] decades ago for somewhat hypothetical, magneto-dielectric particles. Recent theoretical work [16] predicted that real, small dielectric particles made of nonmagnetic materials should present scattering properties similar to those previously reported for magneto-dielectric particles. Specifically, the coherent superposition between the electric and magnetic induced dipoles of non-magnetic particles [16,17] leads to configurations with minimum, or zero, backscattering cross section (the so-called first Kerker condition -FKC-) [15,18,19]. The first experimental confirmations of the FKC have been reported very recently both in the microwave [20] and visible frequency ranges [21,22].

From a practical point of view, due to their unique properties, particles showing zero backscattering with large scattering cross section in forward direction are of great interest in various fields, e.g. as nanoprobes for laser tractor beams [23-26] and other devices based on optically induced "negative forces" [27] or as light diffusing elements for efficiency enhancement in photonic devices. As such, this work contains a search of real materials of which pure dielectric spheres with magneto-dielectric properties can be made with such properties. As we will show, there is an optimal particle refractive index such that nanoparticles present a FKC while they possess a large scattering cross-section. Thus transmitting the maximum diffused energy into the receiving array of storing elements.

We first address spheres with zero absorption, i.e. $Q_{ext}$ = $Q_{sca}$, the extinction cross section being given in terms of the scattering and absorption cross sections as: $Q_{ext} = Q_{sca} + Q_{abs}$. Taking into account the Mie decomposition of the scattered field for a plane wave impinging the dielectric sphere into the series of electric and magnetic Mie coefficients [28], Fig. 1a shows the backscattering efficiency $Q_b$ [28] as a function of wavelength ($\lambda$) for a incident plane wave shining on a spherical particle with radius $a$ =0.24μm . The refractive index of the particle is $m$= 3.25. The spectrum represented by $Q_b$ shows zero or almost zero values within the calculated wavelength range. This means nearly zero backward scattering of the particle at several different wavelengths, such as $\lambda$=1.8μm and $\lambda$=0.80μm, which are indicated by the green and the red solid lines, respectively. The total extinction cross-section ($Q_{ext}$) corresponding to the spectrum of $Q_b$ and the electric ($Q_{ext\_a1}$) and magnetic dipoles ($Q_{ext\_b1}$) in the Mie expansion is shown in Figure 1b.

A comparison between the magnitudes of $Q_{ext}$ at these two selected wavelength shows a much larger scattering cross-section at $\lambda$=1.8μm than at $\lambda$=0.8μm, as depicted by the horizontal dashed line in Figure 1b. In fact, $\lambda$=1.8μm is the target particle's dipolar FKC at which, $Q_{ext\_a1} = Q_{ext\_b1}$ [15,16,19,20].

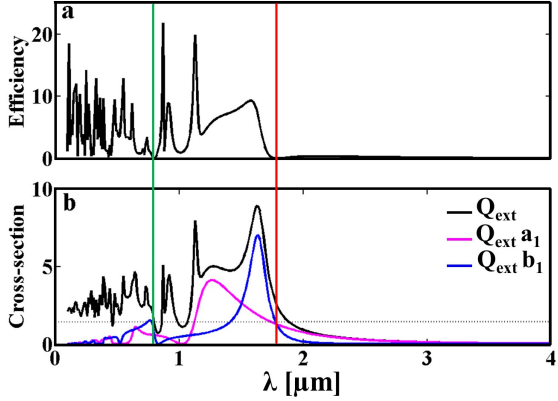

Fig. 1. Backscattering efficiency and extinction cross-section. (a) The efficiency of backscattering ($Q_b$) for a spherical particle with diameter of 0.48 µm and refractive index $m = 3.25$. (b) The total extinction cross-section ($Q_{ext}$) of the particle, and the individual contributions of the electric ($Q_{ext\_a1}$) and magnetic dipoles ($Q_{ext\_b1}$). The red and the green solid curves indicate two wavelength positions where backward scattering is zero.

Figure 2a shows the sphere backscattering efficiency $Q_b$ as a function of its refractive index $m$ and the size parameter $y=2\pi am/\lambda$. The blue regions in the map correspond to zero scattering intensity in the backward direction. The vertical blue vertical band around $y = 2.75$ demonstrates that particles with such size parameter have zero backward scattering for any refractive index $m$ considered. Notice that the zero backscattering condition ($\lambda=1.8$ µm, $m = 3.25$) indicated by the red solid line in Fig. 1 corresponds to $y = 2.72$ and, therefore, it is contained within this blue band. In fact, this blue band shows the places where the FKC holds for particles with refractive index between 2 to 5. This manifests the scaling property of these semiconductor particles, (cf. also Fig.2 of Ref. [17] and Refs. [20,22]), characterized by the straight line:

$$a \approx \frac{y}{2\pi\,m}\lambda = \frac{0.43}{m}\lambda \quad (1)$$

which defines the correlation between particles' radius $a$ and the wavelength of the incident wave $\lambda$ at the FKC when their refractive index $m$ varies.

Even though the backward scattering can also be zero in the regime of larger size parameters, as shown by the other blue regions, the corresponding extinction cross section, as depicted in Fig. 2b, are smaller than the $Q_{ext}$ at the FKC. Figure 2 then shows that, in the range of $m=2$ to $m=5$, the condition given by Eq. (1) correspond to particles having both zero backscattering and a relatively large extinction cross-section.

Our aim in this work is to find the material composing the sphere with the largest scattering cross-section at wavelengths of zero backscattering efficiency, On top of Fig. 3 we plot $Q_{ext}$ versus $m$ for $y$ values within the blue band in Fig. 2a (i.e. where the first FKC holds). Particles with refractive index near 2.47 have the largest extinction cross-section under its FKC, (which takes place at $\lambda = 1.38$ µm for $a = 0.24$µm).

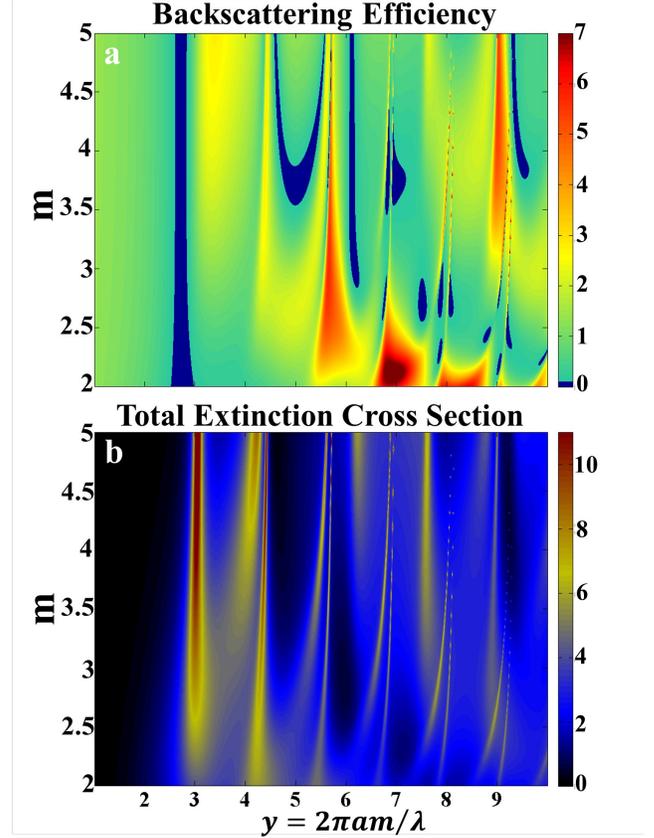

Fig. 2. Color maps of (**a**) Backscattering efficiency ($Q_b$) for spherical non-absorbing particles as a function of their refractive index $m$ and size parameter $y = 2\pi am/\lambda$. Deep blue regions correspond to zero backscattering. (**b**) Corresponding total extinction cross-section ($Q_{ext}$).

Typical backscattering patterns with different scattering cross-sections, computed from the differential scattering cross-section, are illustrated in Figs.3(a)-(c). The three dimensional scattering pattern labeled by (a) in Fig. 3 is due to a particle with refractive index $m=2.5$ (under plane wave illumination inciding from $z= -\infty$), while (b) and (c) correspond to the particle studied in Fig.1 at the two selected wavelengths: $\lambda = 1.8$ µm (at FKC) and $\lambda = 0.80$ µm, respectively. For a better comparison, Fig. 3(d) shows the projection of the three patterns (a), (b), and (c) of the scattered intensity spatial distribution on the $yz$-plane.

Up till now we assumed that all particles were non-absorbing within the considered wavelength range. However, slightly absorbing materials are commonly used in optical devices. Silicon particles, which absorb in the visible, are among those most frequently used in solar cells [13]. Si particles however are not optimal to enhance scattered light at the FKC. For wavelengths larger than ≈ 600nm, where Si absorption is negligible, we can use Eq. 1 to determine the FKC, with $m(\lambda)$ obtained from Ref. [29]. For example, Si nanoparticles with radius 75nm present the FKC at $\lambda$ = 670 nm with m=3.83 which leads to an extinction cross section a factor of 3 smaller than the one obtained for an m=2.5 material (see Fig. 3).

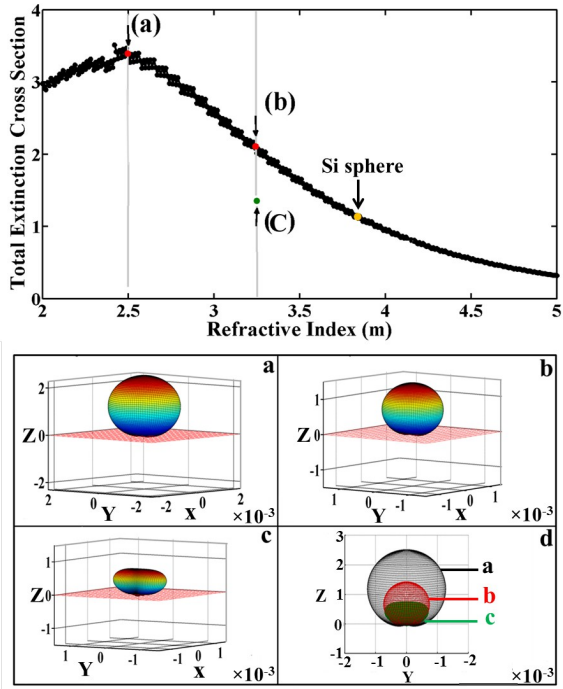

Fig. 3. Top graph: Normalized extinction cross-section $[Q_{ext} / (\pi a^2)]$ at the first Kerker's condition as a function of the refractive index. The yellow dot corresponds to a Si sphere with radius $a=75nm$, illuminated at $\lambda_0 = 640nm$, and $m\lambda_0 = 3.83$. The zero backscattering condition can be observed in the three scattered intensity 3D patterns: (a), (b), and (c), corresponding to the three selected data points: **(a)**, **(b)**, and **(c)** of the top graph, respectively. The inset (d) shows the projection on the *yz*-plane of these three scattered intensity distributions (a), (b), and (c).

It is now appropriate to identify a few examples of real materials having the desired optical response. In this respect we wish to quote a related work [30], of which we became aware after submitting our research to publication.

Fig. 4 shows the correlation between the particle radius *a* and the wavelength of the incident light, i.e. Eq. (1), when the refractive index is in the range of $m = 2.47 \pm 0.1$, which corresponds to the correlation curve shown by the black solid line and the shadow area between the two blue dashed-lines. Three examples of real materials possessing a refractive index within this defined range (cf. [29]), at a variable wavelength, are given in Fig. 4. These three materials are: diamond, titanium dioxide ($TiO_2$), and strontium titanate ($SrTiO_3$), which are represented by the green, red and blue symbols respectively. We should notice that all these examples identified in this letter have negligible absorption at the specified wavelengths [29].

In summary, by using standard Mie theory for lossless dielectric spheres, we have shown that there is an optimum particle refractive index ($m = 2.47$), which yields maximum forward scattering without backwards scattering. At the optimal condition, the total scattering cross section can be as large as $\approx$ 3.5 times the geometrical cross section ($\approx 3.5\ \pi\ a^2$) with almost zero backscattered light.

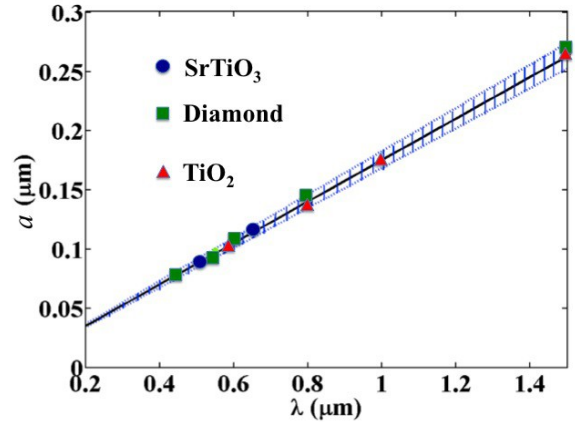

Fig. 4. Correlation between the radius of a spherical particle and the wavelength of the incident plane wave ($\lambda$) at the FKC. The refractive index is in the range of $m = 2.47\pm 0.1$, which correspond to the shadow area between the blue dashed-lines. The size parameter is taken to be the average value $y = 2.75$. Three non-absorbing materials, diamond (green squares), titanium dioxide (TiO2, red triangles) and strontium titanate (SrTiO3, blue dots), are given as real examples at different wavelength

### Acknowledgments

Work supported by the Ministerio de Economia y Competitividad of Spain through grants FIS2012-36113-C03-03 and FIS2014-55563-REDC, and by the Comunidad de Madrid P2009/TIC-1476. J.J.S. acknowledges an Ikerbasque Visiting Fellowship. We thank discussions with Prof. F. Moreno on topics related to this research.